\documentclass[12pt]{article}
\usepackage{amsmath}
\usepackage{graphicx}
\usepackage{amsfonts}
\usepackage{amssymb}
\newcommand{\be}{\begin{equation}}
\newcommand{\ee}{\end{equation}}
\newcommand{\ba}{\begin{eqnarray}}
\newcommand{\ea}{\end{eqnarray}}

\begin{document}

%%%%%%%%%%%%%%%%%%% title page %%%%%%%%%%%%%%%%%%%%%%%%%%%%%%%%%%%%
%
%
%
\begin{titlepage}
\nopagebreak

\begin{flushleft}
CITUSC/02-004\hfill hep-th/0202030\\
UT-983\hfill
\end{flushleft}
\renewcommand{\thefootnote}{\fnsymbol{footnote}}
\vfill
\begin{center}
{\LARGE Associativity Anomaly}\\
{\LARGE in  String Field Theory}
\vskip20mm
\noindent{\large Itzhak Bars${}^a$
and
Yutaka Matsuo${}^b$}
\vskip5mm
${}^a$ {\it CIT-USC Center for Theoretical Physics}
{\it\& Dept. of Physics and  Astronomy}\\
{\it University of Southern California, Los Angeles, CA 90089-2535,
USA}\\
{\tt E-mail: bars@citusc.usc.edu}
\vskip5mm
${}^b$ {\it Department of Physics, University of Tokyo}\\
{\it Hongo 7-3-1, Bunkyo-ku, Tokyo 113-0033, Japan}\\
{\tt E-mail: matsuo@phys.s.u-tokyo.ac.jp}
\end{center}
\vfill
\begin{abstract}
We give a detailed study of the associativity anomaly
in  open string field theory from the viewpoint of
the split string and Moyal formalisms.
The origin of the anomaly is reduced to
the properties of the special infinite size
matrices which relate the conventional open string to
the split string variables, and is intimately related to
midpoint issues. We discuss two steps
to cope with the anomaly. We identify the field subspace
that causes  the anomaly which is related to the existence
of closed string configurations, and
indicate a decomposition of open/closed string sectors.
We then propose a consistent cut off method with a finite number of
string modes that guarantees associativity at every step of any computation.
\end{abstract}
\vfill
\end{titlepage}
%%%%%%%%%%%%%%%%%%%%%%%%%%%%%%%%%%%%%%%%%%%%%%%%%%%%%%%%
%
%
%

\section{Introduction}

Recent developments in vacuum string field theory \cite{r-RSZ}--\cite{r-AGM}
are promising for a description of D-branes and closed strings from the
viewpoint of open string field theory \cite{r-Witten,r-GJ}. In particular, a
simple picture of the stringy solitons emerges as noncommutative solitons of
open string fields.

The algebraic structure of string field theory is greatly simplified by
describing the open string in terms of two halves separated by a midpoint -
the split string formalism \cite{r-Bordes}\cite{r-RSZ}-\cite{r-Bars}. By doing
so, the open string field is regarded as an infinite dimensional matrix.
Furthermore, by transforming to a Fourier space of the odd full string modes
and using some special matrices that naturally emerged in the split string
formalism (the $T,R$ discussed below), Witten's star product is translated
into the standard Moyal product involving the phase space of the even full
string modes \cite{r-Bars}. This establishes an explicit link between open
string field theory and noncommutative geometry in a form which is familiar in
old \cite{r-Moyal} and recent literature \cite{r-DouglasNekrasov}. In this
context, string field theory computations, including the construction of
noncommutative solitons become greatly simplified \cite{r-BarsMatsuo}.

There are, however, some singularities in the split string formalism that
require deeper understanding. In particular, in the description of D-branes
some infinities and zeroes are encountered \cite{GRSZ,r-MT}. So one must learn
how to consistently extract finite quantities from infinite dimensional matrix
calculations or Moyal-star computations that have singular behavior. Related
phenomena were observed long ago \cite{r-HS,r-Strominger,r-Strominger2}, such
as the breakdown of associativity of the star product for certain string field
configurations. Such anomalies typically appear for string fields that
correspond to closed string excitations, such as those that represent
space-time diffeomorphisms.

The breakdown of associativity would have a huge influence on the very
structure of open string field theory. For example, Witten's action would not
enjoy a gauge symmetry in the presence of anomalies. It is therefore important
to know precisely when and how such an anomaly occurs and how it can be treated.

The purpose of this letter is to present a systematic study of such anomalies.
We will show that the associativity anomaly emerges from the very properties
of the infinite dimensional matrices $T,R$ that relate the open full string
degrees of freedom to the split string degrees of freedom, thus clarifying the
origin and the structure of the anomaly. Indeed, we will see that the
Horowitz-Strominger anomaly is hidden in the matrices $T,R$ themselves.

In order to tame such an anomaly, we will discuss two steps: (1) separation of
the open/closed string sectors, (2) a consistent cutoff method.

In the first step, we study the structure of the Hilbert space for split
strings more carefully. We find that the Hilbert space can be decomposed into
two sectors. The first sector is the subspace in which associativity is
maintained. We may regard it as the Hilbert space of open string fields. In
the second sector associativity is explicitly broken. This subspace is
characterized by the fact that under star products with singular fields the
location of the midpoint shifts (contrary to the definition of the original
star product). Thus, we show the simplified origin of the anomaly, with a
direct relation to the Horowitz-Strominger anomaly, through its relation to
the gauge variation of closed string degrees of freedom that are hidden in the
open string formalism.

It is not clear how to precisely separate the open/closed sectors while
maintaining the infinite number of string modes. Therefore, in the second
step, we propose a consistent cutoff method using a finite number of string
modes, and sending the number of modes to infinity at the end of computations.
The essence of our cutoff method is to maintain all the crucial algebraic
relations satisfied by the matrices $T$ and $R$ for any number of modes. This
cutoff method is then valid in both the split string and Moyal formalisms.
With a finite number of modes, associativity is maintained at all stages of
any computation. When the number of modes is sent to infinity the origin of
the anomaly emerges in the form of $\frac{\infty}{\infty}.$ The ambiguity in
such quantities is seen to be the origin of the anomaly. With the consistent
cutoff method the ambiguity is resolved and a unique value is obtained in the
limit. With the cutoff method all quantities of the open string field theory
(off-shell vertex, integration, etc.) are readily expressed in terms of a
finite number of modes, and computations are carried out in a straightforward
way without worrying about the associativity anomaly.

We expect that our consistent cut-off theory would also be quite useful in the
numerical study of vacuum string field theory since it is a more reliable
method as compared to the level truncation which has been used in the recent literature.

\section{Split String and Moyal Formalisms}

We first recall the basic definitions of the split string and Moyal formalisms
in order to set up the notation \cite{r-RSZ}-\cite{r-Bars}. For ease of
notation space-time indices and ghost degrees of freedom will be suppressed in
most formulas.

In Witten's open string field theory, the three string vertex operator is
defined by an overlap of the right half of the first open string with the left
half of the second,
\begin{equation}
(\Psi_{1}\star\Psi_{2})[z(\sigma)]\equiv\int\Psi_{1}[x(\sigma)]\Psi
_{2}[y(\sigma)]\prod_{\pi/2\leq\sigma\leq\pi}\delta\lbrack x(\sigma
)-y(\pi-\sigma)]\,\,dx(\sigma)\,dy(\pi-\sigma)\,\,,
\label{split-string-product}%
\end{equation}
with the identification $z(\sigma)=x(\sigma)$ for $0\leq\sigma\leq\pi/2 $ and
$z(\sigma)=y(\sigma)$ for $\pi/2\leq\sigma\leq\pi$.

The mode expansion of the open string,
\begin{equation}
x(\sigma)=x_{0}+\sqrt{2}\sum_{n=1}^{\infty}x_{n}\cos(n\sigma)
\label{mode-expansion}%
\end{equation}
is not the most convenient set of degrees of freedom to describe the star
product since one cannot say whether $x_{n}$ belongs to the left or right side
of the string. In the operator formalism of the open string field theory, such
description causes the Neumann coefficients appearing in the three string
vertex operator to become rather complicated matrices. This obscures the
understanding of the overall structure and leads to rather complex computations.

The Moyal formulation is obtained by performing a Fourier transform on the odd
string modes. If the original string field written in terms of modes is
$\psi\left(  x_{0},x_{2n},x_{2n-1}\right)  ,$ its Fourier image in the Moyal
basis is $A(\bar{x},x_{2n},p_{2n})$ given as follows \cite{r-Bars}%

\begin{equation}
A(\bar{x},x_{2n},p_{2n})=\det\left(  2T\right)  ^{d/2}\left(  \int
dx_{2n-1}^{\mu}\right)  \,e^{-\frac{2i}{\theta}\eta_{\mu\nu}\sum
_{k,l=1}^{\infty}p_{2k}^{\mu}T_{2k,2l-1}x_{2l-1}^{\nu}}\psi(x_{0}%
,x_{2n},x_{2n-1})\ \label{A}%
\end{equation}
where $d$ is the number of dimensions (26 plus 1 for the bosonized ghosts),
$\theta\,$\ is a parameter that has units of area in phase space,
$T_{2n,2m-1}$ is a special infinite matrix intimately connected to split
strings (see below) and $\bar{x}$ is the string midpoint which may be
rewritten in terms of $x_{0},x_{2n}$ as $\bar{x}=x_{0}+\sqrt{2}\sum
_{n=1}^{\infty}x_{2n}\left(  -1\right)  ^{2n}.$ Then the Witten star product
becomes the Moyal star product in the phase space of each even mode except the
midpoint
\begin{equation}
\left(  A\star B\right)  (\bar{x},x_{2n},p_{2n})=e^{\frac{3i}{2}\bar{x}_{27}%
}A\left(  \bar{x},x_{2n},p_{2n}\right)  \,\,e^{\frac{i\theta}{2}\eta_{\mu\nu
}\sum_{n=1}^{\infty}\left(  \frac{\overleftarrow{\partial}}{\partial
x_{2n}^{\mu}}\frac{\overrightarrow{\partial}}{\partial p_{2n}^{\nu}}%
-\frac{\overleftarrow{\partial}}{\partial p_{2n}^{\nu}}\frac{\overrightarrow
{\partial}}{\partial x_{2n}^{\mu}}\right)  }B\left(  \bar{x},x_{2n}%
,p_{2n}\right)
\end{equation}
Note that the product is local at the midpoint in all dimensions, and that
there is a midpoint insertion $e^{i3\bar{x}^{27}/2}$ in the 27$^{th}$
dimension which is the bosonized ghost coordinate. It is understood that the
midpoint ghost insertion is present in all versions of the star product
although it is not always explicitly indicated. For simplicity of notation we
will continue this tradition of omitting the midpoint insertion in our
formulas below unless we need to do an explicit computation. This
reformulation of the star product greatly simplifies computations of
interacting string fields as shown with many examples in \cite{r-BarsMatsuo}.

The split string formalism defines split string modes which are also
convenient to describe string interactions. In terms of the continuous
parameter $\sigma$, these are defined by explicitly splitting the left and
right variables of the open string relative to a midpoint at $\sigma=\frac
{\pi}{2}$
\begin{equation}
l(\sigma)\equiv x(\sigma),\quad r(\sigma)\equiv x(\pi-\sigma),\quad\mbox
{for}\ \ 0\leq\sigma\leq\pi/2.
\end{equation}
With these new variables, the star product can be written as an infinite
matrix multiplication,
\begin{equation}
(\Psi_{1}\star\Psi_{2})[l(\sigma),r(\sigma)]=\int\prod_{0\leq\sigma\leq\pi
/2}dt(\sigma)\Psi_{1}[l(\sigma),t(\sigma)]\Psi_{2}[t(\sigma),r(\sigma)]
\end{equation}
This expression may be rewritten in terms of the split string modes discussed
below. The open string variable $x(\sigma)$ has no a priori boundary condition
at the midpoint. Therefore, a subtlety in identifying the split string modes
is the boundary condition of the half-string variables $l(\sigma),r(\sigma)$
at the midpoint. Up to this point, two standard choices have been considered,
the Dirichlet and Neumann boundary conditions \cite{r-Bordes,r-Bars}. While we
do not exclude other possibilities, we will concentrate on these two choices
in the following. Either case is compatible with the Moyal basis given above
\cite{r-Bars}.

\subsection{Dirichlet at the midpoint}

We first examine the Dirichlet case $\bar{x}=x\left(  \frac{\pi}{2}\right)
=l(\frac{\pi}{2})=r(\frac{\pi}{2})$. Since we have Neumann boundary conditions
at the other end of $l\left(  \sigma\right)  $ or $r\left(  \sigma\right)  $,
we arrive at the mode expansion in terms of the odd cosines,
\begin{equation}
l(\sigma)=\bar{x}+\sqrt{2}\sum_{n=1}^{\infty}l_{2n-1}\cos(2n-1)\sigma
\,\,,{\quad}r(\sigma)=\bar{x}+\sqrt{2}\sum_{n=1}^{\infty}r_{2n-1}%
\cos(2n-1)\sigma\,\,.
\end{equation}
The Fourier coefficients are related with each other as,
\begin{align*}
l_{2n-1}  &  =\frac{2\sqrt{2}}{\pi}\int_{0}^{\frac{\pi}{2}}d\sigma\,\left(
l\left(  \sigma\right)  -\bar{x}\right)  \cos\left(  2n-1\right)  \sigma
=\frac{2\sqrt{2}}{\pi}\int_{0}^{\frac{\pi}{2}}d\sigma\,\left(  x\left(
\sigma\right)  -\bar{x}\right)  \cos\left(  2n-1\right)  \sigma\\
r_{2n-1}  &  =\frac{2\sqrt{2}}{\pi}\int_{0}^{\frac{\pi}{2}}d\sigma\,\left(
r\left(  \sigma\right)  -\bar{x}\right)  \cos\left(  2n-1\right)  \sigma
=\frac{2\sqrt{2}}{\pi}\int_{0}^{\frac{\pi}{2}}d\sigma\,\left(  x\left(
\pi-\sigma\right)  -\bar{x}\right)  \cos\left(  2n-1\right)  \sigma\\
x_{n\neq0}  &  =\frac{\sqrt{2}}{\pi}\int_{0}^{\pi}d\sigma\,x(\sigma
)\,\cos\left(  n\sigma\right)  =\frac{\sqrt{2}}{\pi}\int_{0}^{\frac{\pi}{2}%
}d\sigma\left[  l\left(  \sigma\right)  +\left(  -1\right)  ^{n}r\left(
\sigma\right)  \right]  \cos\left(  n\sigma\right)  .
\end{align*}
They imply
\begin{align}
x_{2n-1}  &  =\frac{1}{2}\left(  l_{2n-1}-r_{2n-1}\right) \label{lx-odd}\\
x_{2n\neq0}  &  =\frac{1}{2}\sum_{m=1}^{\infty}T_{2n,2m-1}\left(
l_{2m-1}+r_{2m-1}\right) \label{lx-even}\\
x_{0}  &  =\bar{x}+\frac{1}{2\sqrt{2}}\sum_{m=1}^{\infty}T_{0,2m-1}\left(
l_{2m-1}+r_{2m-1}\right)  \label{xbarx-odd}%
\end{align}
where
\begin{align}
T_{2n,2m-1}  &  =\frac{4}{\pi}\int_{0}^{\frac{\pi}{2}}d\sigma\cos\left(
\left(  2n\right)  \sigma\right)  \cos\left(  \left(  2m-1\right)
\sigma\right) \nonumber\\
&  =\frac{2\left(  -1\right)  ^{m+n+1}}{\pi}\left(  \frac{1}{2m-1+2n}+\frac
{1}{2m-1-2n}\right)  . \label{T}%
\end{align}
This matrix $T$ is directly related to the matrix $X$ in \cite{r-GJ,r-GT} as
follows,
\begin{align}
X_{2m-1,2n}  &  =-X_{2n,2m-1}=iT_{2n,2m-1},\quad(n>0)\\
X_{0,2m-1}  &  =\frac{i}{\sqrt{2}}T_{0,2m-1}%
\end{align}
The inverse relations of Eqs.(\ref{lx-odd}--\ref{xbarx-odd}) are
\begin{align}
l_{2m-1}  &  =x_{2m-1}+\sum_{n=1}^{\infty}R_{2m-1,2n}x_{2n}\label{l}\\
\bar{x}  &  =x_{0}+\sqrt{2}\sum_{n=1}^{\infty}\left(  -1\right)  ^{n}%
x_{2n}\label{mid}\\
r_{2m-1}  &  =-x_{2m-1}+\sum_{n=1}^{\infty}R_{2m-1,2n}x_{2n} \label{r}%
\end{align}
where
\begin{align}
R_{2m-1,2n}  &  =\frac{4}{\pi}\int_{0}^{\frac{\pi}{2}}d\sigma\cos\left(
2m-1\right)  \sigma\,\left[  \cos2n\sigma-\left(  -1\right)  ^{n}\right] \\
&  =\frac{4n\left(  -1\right)  ^{n+m}}{\pi\left(  2m-1\right)  }\left(
\frac{1}{2m-1+2n}-\frac{1}{2m-1-2n}\right)  .
\end{align}
Note that
\begin{equation}
R_{2m-1,2n}=T_{2n,2m-1}\frac{\left(  2n\right)  ^{2}}{\left(  2m-1\right)
^{2}}=T_{2n,2m-1}-\left(  -1\right)  ^{n}T_{0,2m-1} \label{RT}%
\end{equation}
It must be mentioned that $R_{2k-1,2m}$ is the inverse of $T_{2m,2n-1}$ on
both sides
\begin{equation}
(RT)_{2m-1,2k-1}=\delta_{m,k},\qquad(TR)_{2m,2k}=\delta_{m,k}. \label{inverse}%
\end{equation}
From Eqs.(\ref{RT}) and (\ref{inverse}) one obtains the relations
\begin{align}
\sum_{n=1}^{\infty}T_{2n,2m-1}\left(  2n\right)  ^{2}T_{2n,2k-1}  &  =\left(
2m-1\right)  ^{2}\delta_{m,k}\label{eig1}\\
\sum_{n=1}^{\infty}T_{2m,2n-1}\frac{1}{\left(  2n-1\right)  ^{2}}T_{2k,2n-1}
&  =\frac{1}{\left(  2m\right)  ^{2}}\delta_{m,k}\\
\sum_{n=1}^{\infty}R_{2n-1,2m}\left(  2n-1\right)  ^{2}R_{2n-1,2k}  &
=\left(  2m\right)  ^{2}\delta_{m,k}\\
\sum_{n=1}^{\infty}R_{2m-1,2n}\frac{1}{\left(  2n\right)  ^{2}}R_{2k-1,2n}  &
=\frac{1}{\left(  2m-1\right)  ^{2}}\delta_{m,k} \label{eig2}%
\end{align}
These equations reflect the fact that the matrices $T$ and $R$ are
transformations between two bases of the form $\cos(2n\sigma),$ $\cos(\left(
2n-1\right)  \sigma)$ which diagonalize the Laplacian, $-\partial_{\sigma}%
^{2}$ with two different boundary conditions.

\subsection{Neumann at the midpoint}

First we note the following properties of trigonometric functions when
$0\leq\sigma\leq{\pi}$ for integers $m,n\geq1$
\begin{align}
\cos((2n-1)\sigma)  &  =sign(\frac{\pi}{2}-\sigma)\sum_{m=1}^{\infty}%
[\cos(2m\sigma)-(-1)^{m}]\ T_{2m,2n-1}\label{trig1}\\
\lbrack\cos(2m\sigma)-(-1)^{m}]  &  =sign(\frac{\pi}{2}-\sigma)\sum
_{n=1}^{\infty}\ \cos((2n-1)\sigma)\ R_{2n-1,2m}. \label{trig2}%
\end{align}
Both sides of these equations satisfy Neumann boundary conditions at
$\sigma=0$ and Dirichlet boundary conditions at $\sigma=\frac{\pi}{2}$, and
both are equivalent complete sets of trigonometric functions for the range
$0\leq\sigma\leq\frac{\pi}{2}$ . In the previous section we made the choice of
expanding $l(\sigma),r(\sigma)$ in terms of the odd modes. Now we see that we
could also expand them in terms of the even modes as follows
\begin{equation}
l(\sigma)=\bar{x}+\sqrt{2}\sum_{m=1}^{\infty}\ l_{2m}[\cos(2m\sigma
)-(-1)^{m}]=l_{0}+\sqrt{2}\sum_{m=1}^{\infty}\ l_{2m}\cos(2m\sigma)
\label{evenmodes}%
\end{equation}
and similarly for $r(\sigma)$. The even modes $l_{2m}$ are now associated with
$\cos(2m\sigma)$ which is a complete set that satisfies Neumann boundary
conditions at $\sigma=0,\frac{\pi}{2}.$ Comparing to the expressions in the
previous subsection, and using (\ref{trig1},\ref{trig2}) we can find the
relation between the odd modes $(l_{2n-1},r_{2n-1})$ and the even modes
$(l_{2n},r_{2n})$
\begin{align}
l_{2n-1}  &  =\sum_{m=1}^{\infty}\ R_{2n-1,2m}l_{2m},\ \ \ \ l_{2m}=\sum
_{n=1}^{\infty}\ T_{2m,2n-1}l_{2n-1}\label{evenodd1}\\
r_{2n-1}  &  =\sum_{m=1}^{\infty}\ R_{2n-1,2m}r_{2m},\ \ \ \ r_{2m}=\sum
_{n=1}^{\infty}\ T_{2m,2n-1}r_{2n-1} \label{evenodd2}%
\end{align}
Furthermore, by using the relation between the odd string modes $(l_{2n-1}%
,\bar{x},r_{2n-1})$ and the full string modes $(x_{0},x_{2n},x_{2n-1})$ in
Eqs.(\ref{l}-\ref{r}) or by direct comparison to $x(\sigma)$, we derive the
relation between the even split string modes and the full string modes.
\begin{align}
l_{2m}  &  =x_{2m}+\sum_{n=1}^{\infty}T_{2m,2n-1}x_{2n-1},\qquad r_{2m}%
=x_{2m}-\sum_{n=1}^{\infty}T_{2m,2n-1}x_{2n-1},\label{even}\\
l_{0}  &  =x_{0}+\frac{1}{\sqrt{2}}\sum_{n=1}^{\infty}T_{0,2n-1}%
x_{2n-1},\qquad r_{0}=x_{0}-\frac{1}{\sqrt{2}}\sum_{n=1}^{\infty}%
T_{0,2n-1}x_{2n-1}.. \label{l0r0}%
\end{align}
The inverse relation is
\begin{equation}
x_{0}=\frac{l_{0}+r_{0}}{2},\quad x_{2m}=\frac{l_{2m}+r_{2m}}{2},\qquad
x_{2m-1}=\sum_{n=1}^{\infty}R_{2m-1,2n}\frac{l_{2n}-r_{2n}}{2}. \label{evenn}%
\end{equation}
Note that the matching condition at the midpoint $l(\pi/2)=r(\pi
/2)=x(\pi/2)=\bar{x}$ is satisfied by the even modes. This is evident from the
first expression in Eq.(\ref{evenmodes}) and also by noting that $l_{0}-r_{0}$
automatically obeys the relation
\begin{equation}
l_{0}-r_{0}=\sqrt{2}\sum_{n=1}^{\infty}T_{0,2n-1}x_{2n-1}=-\sqrt{2}\sum
_{n=1}^{\infty}(l_{2n}-r_{2n})\left(  -1\right)  ^{n} \label{l0-r0}%
\end{equation}
as seen by using Eqs.(\ref{l0r0},\ref{evenn}) and inserting the relation
$\bar{v}R=\bar{w}$ given below in Eq.(\ref{v}). In working purely with even
split string modes, Eq.(\ref{l0-r0}) is a constraint on $\left(  l_{0,}%
,r_{0},l_{2n},r_{2n}\right)  $ that must be imposed among those modes.
However, an alternative strategy is to use the unconstrained modes $\left(
\bar{x},l_{2n},r_{2n}\right)  $ as the independent modes instead of the
constrained modes $\left(  l_{0,},r_{0},l_{2n},r_{2n}\right)  .$ In this case,
instead of Eq.(\ref{evenn}), the center of mass $x_{0}$ is given in terms of
the split string modes $\left(  \bar{x},l_{2n},r_{2n}\right)  $ by
\begin{equation}
x_{0}=\bar{x}-\sqrt{2}\sum_{n=1}^{\infty}\frac{l_{2m}+r_{2m}}{2}\left(
-1\right)  ^{n},
\end{equation}
while the expression for $l_{0}-r_{0}$ never enters and can take its allowed
values in terms of $\left(  l_{2n}-r_{2n}\right)  $ as seen in Eq.(\ref{l0-r0}).

\subsection{Relations among $T,R,v,w$}

More relations among the special matrices $T,R$ can be compactly written in
matrix notation by defining the even and odd vectors $w,v$
\begin{equation}
\quad w_{2m}=\sqrt{2}\left(  -1\right)  ^{m+1},\quad v_{2n-1}=\frac{1}%
{\sqrt{2}}T_{0,2n-1}=\frac{2\sqrt{2}}{\pi}\frac{\left(  -1\right)  ^{n+1}%
}{2n-1}, \label{vandw}%
\end{equation}
and then noting the following identities among these matrices,
\begin{align}
TR  &  =1,\quad RT=1,\quad R=\bar{T}+v\bar{w},\quad R=\kappa_{o}^{-2}\bar
{T}\kappa_{e}^{2},\label{tr}\\
v  &  =\bar{T}w,\quad w=\bar{R}v,\;\quad\bar{R}R=1+w\bar{w},\quad\bar
{T}T=1-v\bar{v}\label{v}\\
T\bar{T}  &  =1,\quad Tv=0,\quad\bar{v}v=1. \label{tt2}%
\end{align}
where the bar on a symbol means transpose of the matrix. In Eq.(\ref{tr}) we
have defined the odd and even diagonal matrices
\begin{equation}
\kappa_{o}=diag\left(  2n-1\right)  ,\quad\kappa_{e}=diag\left(  2n\right)  ,
\label{kappas}%
\end{equation}
to reproduce Eq.(\ref{RT}). We recall that the meaning of the eigenvalues of
$\kappa_{o},\kappa_{e}$ are the frequencies of oscillation of the string modes.

As we see in the next section, these identities, while they come from the
absolutely convergent sums, are not consistent with each other in the sense
that they break associativity when some of these matrices occur in double
sums. The culprits are the relations in Eq.(\ref{tt2}) and the underlying
reason is the infinite norm $\bar{w}w=\infty.$ In the final section, we
propose a finite size version of the matrices $T$, $R$, $w$ and $v$ to make
all matrix relations consistent with associativity. Here we give a simple
sketch of our idea. We suppose that we have a regularization scheme where a
suitably redefined $w$ has a finite norm ($\bar{w}w=$finite). Then there is a
unique way to impose associativity consistent with the definitions of
$T,R,v,w$ as expressed in Eqs.(\ref{tr}, \ref{v}). Associativity forces us to
modify the formulas in (\ref{tt2}) to the unique form
\begin{align}
T\bar{T}  &  =1-\frac{w\bar{w}}{1+\bar{w}w},\quad Tv=\frac{w}{1+\bar{w}%
w},\quad\bar{v}v=\frac{\bar{w}w}{1+\bar{w}w},\label{tt}\\
Rw  &  =v(1+\bar{w}w),\quad R\bar{R}=1-v\bar{v}\left(
1+\bar{w}w\right)  .
\label{Rw}%
\end{align}
One derives them as, for example, $Tv=T(\bar{T}w)=T(R-v\bar{w})w=TRw-Tv\bar
{w}w=w-(Tv)(\bar{w}w)$, which implies $Tv=\frac{w}{1+\bar{w}w}$. Of course, in
the infinite norm limit of $w$, one reproduces (\ref{tt2}). We will often come
back to this issue in the text. The details of the cutoff procedure with
finite rank matrices that preserve all the relations above are presented in
section 5.

\section{Associativity anomaly}

In this section, we explain the appearance of the associativity
anomaly hidden in the split string formalism. The matrix algebras
between $T$,$R$, $w $, $v$ are defined by the absolutely
convergent infinite sums as emphasized above. However the
\emph{double} sum appearing in the product of three elements can
be only conditionally convergent and the two infinite sums in
different order do not in general give the same answer, thus
producing an anomaly. We will see that physically the anomaly
appears as the subtleties at the midpoint.

We first show the most typical example. The matrices $T_{2n,2m-1}$ and
$v_{2m-1}=\frac{1}{\sqrt{2}}T_{0,2m-1}$ defined by Eq.(\ref{T}) satisfy
$v=\bar{T}w,$ or
\begin{equation}
T_{0,2n-1}=-2\sum_{k=1}^{\infty}\left(  -1\right)  ^{k}T_{2k,2n-1}. \label{T0}%
\end{equation}
Going back to the original definition in terms of the integrals of cosines as
in Eq.(\ref{T}), this equation is satisfied as follows%
\begin{equation}
\int_{0}^{\pi/2}\cos((2n-1)\sigma)\,d\sigma=-2\sum_{k=1}^{\infty}\left(
-1\right)  ^{k}\int_{0}^{\pi/2}\cos((2n-1)\sigma)\,\cos\left(  2k\sigma
\right)  \,d\sigma
\end{equation}
But this is rewritten as
\begin{equation}
\int_{0}^{\pi/2}\cos((2n-1)\sigma)\,\delta(\sigma-\pi/2)\,d\sigma=0,
\label{midpoint-delta}%
\end{equation}
where the periodic delta function is given by%
\begin{align}
\delta(\sigma-\pi/2)  &  =\frac{1}{\pi}-\frac{2}{\pi}\sum_{k=1}^{\infty
}\left(  -1\right)  ^{k}\cos(2k\sigma)\nonumber\\
&  =\frac{1}{\pi}+\sqrt{2}\sum_{n=1}^{\infty}\frac{-w_{2n}}{\pi}\cos\left(
2n\sigma\right)  \label{delta}%
\end{align}
Thus, through the delta function we see that $v=\bar{T}w$ is a relation
involving the midpoint. Together with the identities $T\bar{T}=1,$ $Tv=0,$
$\bar{v}v=1$ given in Eq.(\ref{tt2}), these matrices display an associativity
anomaly as follows
\begin{align}
&  (T\bar{T})w=1\cdot w=w\quad\mbox{versus}\quad T(\bar{T}w)=Tv=0,\\
&  (\bar{v}\bar{T})w=0\cdot w=0\quad\mbox{versus}\quad\bar{v}(\bar{T}%
w)=\bar{v}v=1.
\end{align}
Namely $(T\bar{T})w\neq T(\bar{T}w)$ and $(\bar{v}\bar{T})w\neq\bar{v}(\bar
{T}w)$. These examples clearly show the anomaly is intimately related to the midpoint.

Before we move on, let us point out what would happen to the double infinite
sums if the infinite norm $\bar{w}w=\infty$ is not imposed in the single sums,
as would be the case in any cutoff procedure. Then, instead of Eq.(\ref{tt2})
we use Eq.(\ref{tt}). This gives
\begin{equation}
T\bar{T}w=w\frac{\bar{w}w}{1+\bar{w}w},\quad\bar{v}\bar{T}w=\frac{\bar{w}%
w}{1+\bar{w}w}%
\end{equation}
independent of the order of the sums. The anomaly is circumvented if $\bar
{w}w=\infty$ is imposed at the end of the computation since then there is a
unique answer. After emphasizing the significance of the anomaly in terms of
midpoint issues, we will propose a consistent cutoff procedure that will rely
on this observation.

In the following, we show more specifically how the anomaly arises for the two
choices of the midpoint boundary conditions considered in the previous section.

\subsection{Dirichlet at the midpoint (odd modes)}

We write the relation between the full string modes and the split string modes
(\ref{lx-odd}--\ref{xbarx-odd},\ref{l}--\ref{r}) in matrix notation
\begin{align}
\left(
\begin{array}
[c]{c}%
x_{0}\\
x^{(e)}\\
x^{\left(  o\right)  }%
\end{array}
\right)   &  =\left(
\begin{array}
[c]{ccc}%
1 & \frac{1}{2}\bar{v} & \frac{1}{2}\bar{v}\\
0 & \frac{1}{2}T & \frac{1}{2}T\\
0 & \frac{1}{2} & -\frac{1}{2}%
\end{array}
\right)  \left(
\begin{array}
[c]{c}%
\bar{x}\\
l^{\left(  o\right)  }\\
r^{\left(  o\right)  }%
\end{array}
\right)  \equiv\mathcal{T}^{\left(  o\right)  }\left(
\begin{array}
[c]{c}%
\bar{x}\\
l^{\left(  o\right)  }\\
r^{\left(  o\right)  }%
\end{array}
\right) \label{xy-odd}\\
\left(
\begin{array}
[c]{c}%
\bar{x}\\
l^{\left(  o\right)  }\\
r^{\left(  o\right)  }%
\end{array}
\right)   &  =\left(
\begin{array}
[c]{ccc}%
1 & -\bar{w} & 0\\
0 & R & 1\\
0 & R & -1
\end{array}
\right)  \left(
\begin{array}
[c]{c}%
x_{0}\\
x^{(e)}\\
x^{\left(  o\right)  }%
\end{array}
\right)  \equiv\mathcal{R}^{\left(  o\right)  }\left(
\begin{array}
[c]{c}%
x_{0}\\
x^{(e)}\\
x^{\left(  o\right)  }%
\end{array}
\right)  \label{yx-odd}%
\end{align}
where we use the notation $e$=even, $o$=odd and the right hand sides define
the matrices $\mathcal{T}^{\left(  o\right)  }$ and $\mathcal{R}^{\left(
o\right)  }$. One may check $\mathcal{T}^{\left(  o\right)  }\mathcal{R}%
^{\left(  o\right)  }=\mathcal{R}^{\left(  o\right)  }\mathcal{T}^{\left(
o\right)  }=1$ by using the formulae, $TR=RT=1$, $\bar{v}R=\bar{w}$ and
$\bar{v}=\bar{w}T$. A subtle point in this correspondence is that
$\mathcal{T}^{\left(  o\right)  }$ has a state with zero eigenvalue given by
$\left(  \bar{x},l^{\left(  o\right)  },r^{\left(  o\right)  }\right)
\sim\left(  -1,v,v\right)  \equiv\mathcal{V}^{\left(  o\right)  }$%
\begin{equation}
\mathcal{T}^{\left(  o\right)  }\mathcal{V}^{\left(  o\right)  }\equiv\left(
\begin{array}
[c]{ccc}%
1 & \frac{1}{2}\bar{v} & \frac{1}{2}\bar{v}\\
0 & \frac{1}{2}T & \frac{1}{2}T\\
0 & \frac{1}{2} & -\frac{1}{2}%
\end{array}
\right)  \left(
\begin{array}
[c]{c}%
-1\\
v\\
v
\end{array}
\right)  =\left(
\begin{array}
[c]{c}%
-1+\bar{v}v\\
Tv\\
0
\end{array}
\right)  =0, \label{calV}%
\end{equation}
which follows from $Tv=0$, $\bar{v}v=1$. Note that the eigenstate
$\mathcal{V}^{\left(  o\right)  }$ has finite norm. These facts imply that
associativity is broken explicitly as follows
\begin{equation}
(\mathcal{R}^{\left(  o\right)  }\mathcal{T}^{\left(  o\right)  }%
)\mathcal{V}^{\left(  o\right)  }=\mathcal{V}^{\left(  o\right)  },\quad
\mbox{versus}\quad\mathcal{R}^{\left(  o\right)  }\left(  \mathcal{T}^{\left(
o\right)  }\mathcal{V}^{\left(  o\right)  }\right)  =0.
\end{equation}

The interpretation of the eigenvector $\mathcal{V}^{\left(  o\right)  }$\ is
that the infinitesimal translation of the split string modes given by two
translation parameters $a^{\mu},b^{\mu}$
\begin{equation}
\delta\bar{x}_{\mu}=a_{\mu},\quad\delta l_{2n-1}^{\mu}=b^{\mu}v_{2n-1}%
,\quad\delta r_{2n-1}^{\mu}=b^{\mu}v_{2n-1}, \label{ab1}%
\end{equation}
does not generate any translation of the full string modes $\left(
x_{2n},x_{2n-1}\right)  $ while $x_{0}$ is translated only by the sum $a^{\mu
}+b^{\mu}$ but not the difference $a^{\mu}-b^{\mu}$
\begin{equation}
\delta x_{0}^{\mu}=a^{\mu}+b^{\mu} \label{ab2}%
\end{equation}
So, there is an extra zero mode in the split string formalism as compared to
the full string formalism. In this sense, the correspondence between the split
string modes and the full string modes does not seem to be one-to-one and
either $\bar{x}$ or the variation of $l^{\left(  o\right)  },r^{\left(
o\right)  }$ along $v$ appear to contain an extra zero mode. This redundancy
gives the origin of the anomaly in this case. We will further clarify below
the relation of this anomaly to the Horowitz-Strominger anomaly
\cite{r-HS,r-Strominger,r-Strominger2}, and to the pure midpoint-ghost BRST
operator recently suggested in the vacuum string field theory formalism
\cite{GRSZ}.

As above, in a cutoff scheme, if the infinite norm $\bar{w}w=\infty$ is not
imposed temporarily in the single sums, and we use Eq.(\ref{tt}) instead of
Eq.(\ref{tt2}), we get the temporarily non-zero result
\begin{equation}
\mathcal{T}^{\left(  o\right)  }\mathcal{V}^{\left(  o\right)  }%
\mathcal{=}\left(
\begin{array}
[c]{c}%
-1\\
w\\
0
\end{array}
\right)  \frac{1}{1+\bar{w}w}%
\end{equation}
Then $\mathcal{R}^{\left(  o\right)  }\mathcal{T}^{\left(  o\right)
}\mathcal{V}^{\left(  o\right)  }\mathcal{=V}^{\left(  o\right)  }$ follows
without associativity anomalies in the double sums, provided the infinite norm
$\bar{w}w=\infty$ is not imposed until the end of the computation.

\subsection{Neumann at the midpoint (even modes)}

The relations similar to Eqs.(\ref{xy-odd},\ref{yx-odd}) are,
\begin{align}
\left(
\begin{array}
[c]{c}%
x_{0}\\
x^{(e)}\\
x^{(o)}%
\end{array}
\right)   &  =\left(
\begin{array}
[c]{ccc}%
1 & \frac{1}{2}\bar{w} & \frac{1}{2}\bar{w}\\
0 & \frac{1}{2} & \frac{1}{2}\\
0 & \frac{1}{2}R & -\frac{1}{2}R
\end{array}
\right)  \left(
\begin{array}
[c]{c}%
\bar{x}\\
l^{\left(  e\right)  }\\
r^{\left(  e\right)  }%
\end{array}
\right)  \equiv\mathcal{R}^{\left(  e\right)  }\left(
\begin{array}
[c]{c}%
\bar{x}\\
l^{\left(  e\right)  }\\
r^{\left(  e\right)  }%
\end{array}
\right) \\
\left(
\begin{array}
[c]{c}%
\bar{x}\\
l^{\left(  e\right)  }\\
r^{\left(  e\right)  }%
\end{array}
\right)   &  =\left(
\begin{array}
[c]{ccc}%
1 & -\bar{w} & 0\\
0 & 1 & T\\
0 & 1 & -T
\end{array}
\right)  \left(
\begin{array}
[c]{c}%
x_{0}\\
x^{(e)}\\
x^{(o)}%
\end{array}
\right)  \equiv\mathcal{T}^{\left(  e\right)  }\left(
\begin{array}
[c]{c}%
x_{0}\\
x^{(e)}\\
x^{(o)}%
\end{array}
\right)
\end{align}
There is an eigenvector with zero eigenvalue when $\left(  x_{0}%
,x^{(e)},x^{(o)}\right)  \sim\left(  0,0,v\right)  \equiv\mathcal{V}^{\left(
e\right)  }$%
\begin{equation}
\mathcal{T}^{\left(  e\right)  }\mathcal{V}^{\left(  e\right)  }\equiv\left(
\begin{array}
[c]{ccc}%
1 & -\bar{w} & 0\\
0 & 1 & T\\
0 & 1 & -T
\end{array}
\right)  \left(
\begin{array}
[c]{c}%
0\\
0\\
v
\end{array}
\right)  =\left(
\begin{array}
[c]{c}%
0\\
Tv\\
-Tv
\end{array}
\right)  =0
\end{equation}
which follows from the single sum in $Tv=0.$ Again, we meet the associativity
anomaly in the double sums
\begin{equation}
\left(  \mathcal{R}^{\left(  e\right)  }\mathcal{T}^{\left(  e\right)
}\right)  \mathcal{V}^{\left(  e\right)  }=\mathcal{V}^{\left(  e\right)
},\quad\mbox{versus}\quad\mathcal{R}^{\left(  e\right)  }\left(
\mathcal{T}^{\left(  e\right)  }\mathcal{V}^{\left(  e\right)  }\right)  =0.
\end{equation}
In this case, we have to be more careful since the zero eigenstate occurs on
the full string side. That is, the translation of the full string mode
$x^{\left(  o\right)  \mu}$ by $\epsilon^{\mu}v$ does not seem to induce any
translation in the split string variables $\left(  \bar{x},l^{\left(
e\right)  },r^{\left(  e\right)  }\right)  $%
\begin{equation}
\left(
\begin{array}
[c]{c}%
\delta\bar{x}\\
\delta l^{\left(  e\right)  }\\
\delta r^{\left(  e\right)  }%
\end{array}
\right)  =\left(
\begin{array}
[c]{ccc}%
1 & -\bar{w} & 0\\
0 & 1 & T\\
0 & 1 & -T
\end{array}
\right)  \left(
\begin{array}
[c]{c}%
0\\
0\\
v
\end{array}
\right)  =\epsilon^{\mu}\left(
\begin{array}
[c]{c}%
0\\
Tv\\
-Tv
\end{array}
\right)  =0
\end{equation}
In this case, the split string modes we have chosen do not seem to be enough
to describe the open string degrees of freedom. However, let us analyze the
zero mode $\left(  l_{0}-r_{0}\right)  $ as given in Eq.(\ref{l0-r0}). From
the expression $l_{0}-r_{0}=2\bar{v}x^{\left(  o\right)  }$ we see that it
certainly translates when the full string mode $x^{\left(  o\right)  \mu}$ is
translated by $\epsilon^{\mu}v,$ that is $\delta\left(  l_{0}^{\mu}-r_{0}%
^{\mu}\right)  =2\epsilon^{\mu}$ after using $\bar{v}v=1.$ This shows that the
infinite sum $\bar{w}\left(  l^{e}-r^{e}\right)  $ also must translate by the
same amount even though the individual $l^{e},r^{e}$ did not seem to
translate
\begin{equation}
\delta\left(  l_{0}^{\mu}-r_{0}^{\mu}\right)  =2\bar{v}\delta x^{\left(
o\right)  }=\bar{w}\left(  \delta l^{e}-\delta r^{e}\right)  =2\epsilon^{\mu}.
\end{equation}
Thus, we see again that double infinite sums $\bar{w}Tv$ must be evaluated
carefully as they are afflicted with the associativity anomaly. Once more, in
a regularized theory, if we use $Tv=w\left(  1+\bar{w}w\right)  ^{-1}$ as in
Eq.(\ref{tt}) instead of the zero value in Eq.(\ref{tt2}), then the correct
result $\bar{w}\left(  \delta l^{e}-\delta r^{e}\right)  =2\epsilon^{\mu}$ is
recovered by setting $\bar{w}w=\infty$ at the end of the calculation.

\subsection{Relation to the Horowitz-Strominger anomaly}

Actually the associativity anomaly which we encountered in this section is the
split string version of the Horowitz-Strominger's anomaly in \cite{r-HS}. In
that paper, the space-time translation generator is represented as the inner
derivative of the open string fields. The generator is represented by the
string field
\begin{equation}
P_{L}|\mathcal{I}\rangle
\end{equation}
where $P_{L}$ is the momentum density integrated over the left half of the
open string. The string configuration described by $P_{L}$ shifts the center
of mass of the full string under commutation using the star product. This
singular behavior gives rise to the Horowitz-Strominger anomaly.

More explicitly, in terms of the vertex operator, they observed that
\begin{align}
(P_{1R}+P_{2L})|V_{1234}\rangle &  =0\label{m-consv}\\
(\bar{x}_{1}-\bar{x}_{3})|V_{1234}\rangle &  =0\\
\lbrack P_{1R}+P_{2L},\bar{x}_{1}-\bar{x}_{3}]  &  =-i/2.
\end{align}
The first equation represents the conservation of the momentum for the four
string interaction. The second represents that the midpoint is fixed for the
interaction. The third equation, however, says that $P$ and $\bar{x}$ does not
commute. Obviously, these equations are not consistent with each other if
associativity is assumed
\begin{equation}
\bar{x}_{1}(P_{1R}|V_{1234}\rangle)- P_{1R}(\bar{x}_{1}|V_{1234}\rangle
)\neq(\bar{x}_{1}P_{1R}-P_{1R}\bar{x}_{1})|V_{1234}\rangle\,\,.
\end{equation}

In the split string formalism, the product is defined by the path integral
over the half string (\ref{split-string-product}).
%\begin{equation}\label{star}
% (\Phi\star\Psi)[l(\sigma),\bar{x},r(\sigma)]= \int [Dt(\sigma)] \Phi[l(\sigma)
%,\bar{x},t(\sigma)]
%\Psi[t(\sigma),\bar{x},r(\sigma)].
%\end{equation}
%
%
%
The momentum conservation (\ref{m-consv}) is represented as the invariance of
the constant shift of the integration variable $t(\sigma)$ on the right hand
side of (\ref{split-string-product}). In this sense, $P_{L,R}$ operator should
induce the infinitesimal translation of $l(\sigma),r(\sigma)$ by a constant.
In the odd modding split string formalism, it is generated by the operator,
\begin{equation}
P_{L}^{\mu}=\sum_{n=1}^{\infty}v_{2n-1}\partial_{l_{2n-1}}^{\mu}\,,\quad
P_{R}^{\mu}=\sum_{n=1}^{\infty}v_{2n-1}\partial_{r_{2n-1}}^{\mu}.
\end{equation}
The sum generates exactly the type of the translation $b_{\mu}$ in
Eq.(\ref{ab1}) which caused the associativity anomaly in our case ($\delta
l_{\mu}^{\left(  o\right)  }=b_{\mu}v=\delta r_{\mu}^{\left(  o\right)  }$).
The associativity anomaly appears there because there is a redundancy in the
split string description. We also noted in Eq.(\ref{ab2}) that this
translation causes a shift in the center of mass coordinate, as claimed by
Horowitz and Strominger. We have therefore identified the Horowitz and
Strominger anomaly with the anomaly in the very matrices $R,T,v,w$ that occur
naturally in the split string formulation.

In the Moyal formulation $-i\partial_{l_{2n-1}}^{\mu}\psi$ corresponds to left
multiplication under the Moyal star product $\sum_{m}\bar{T}_{2n-1,2m}\left(
p_{2m}^{\mu}\star A\right)  $ and $i\partial_{r_{2n-1}}^{\mu}\psi$ corresponds
to right multiplication\footnote{These will be discussed in detail in a future
paper \cite{r-BarsMatsuo}.} $\bar{T}_{2n-1,2m}\left(  A\star p_{2m}^{\mu
}\right)  .$ In particular the sum $\left(  P_{L}^{\mu}+P_{R}^{\mu}\right)
A\left(  \bar{x},x_{2n},p_{2n}\right)  $ is given by the commutator
$i\sum_{n,m}v_{2n-1}\bar{T}_{2n-1,2m}\left(  p_{2m}^{\mu}\star A-A\star
p_{2m}^{\mu}\right)  $. Taking into account $Tv=0,$ we see that the
translation $\left(  P_{L}^{\mu}+P_{R}^{\mu}\right)  A\left(  \bar{x}%
,x_{2n},p_{2n}\right)  $ vanishes unless the string field $A$ is such that the
commutator $\left(  p_{2m}^{\mu}\star A-A\star p_{2m}^{\mu}\right)  $ behaves
like $w_{2m}$ (since the double sum $wTv$ is ambiguous by the anomaly). Such a
string field configuration must involve $\sum\left(  -1\right)  ^{n}x_{2n}$
which is precisely related to the difference between the center of mass and
midpoint $\left(  x_{0}-\bar{x}\right)  $ as in Eq.(\ref{mid}). Hence
Strominger's anomaly is closely connected to the associativity anomaly among
the matrices $R,T,v,w.$

If we follow the discussion in this section, the anomaly would not exist if we
exclude string fields that are non-trivial under the variation induced by
$P_{L}+P_{R}$. If one takes such an approach the excluded string field
configurations would live outside of the open string Hilbert space, and would
belong to the closed string sector that are nontrivial under space-time
diffeomorphisms generated by $P_{L}^{\mu}$ +$P_{R}^{\mu}$ as advocated by Strominger.

\section{Controlling the anomaly}

There are basically two natural ways to control the associativity anomaly. One
method is to use projectors that separate the anomalous sector in the Hilbert
space, thereby separating the open/closed string field sectors. This is along
the lines of an old proposal by Strominger as described below. The other
method is to consider a regularization which is by definition free from
anomaly. In this section we consider the first strategy in the presence of an
infinite number of modes. We first discuss a projection and its relation to
old works. We then point out the relevance to midpoint issues that arise in
recent proposals in the context of vacuum string field theory. In section 5 we
propose another way of controlling the anomaly through a new consistent
regularization using a finite number of modes $N,$ with the cutoff $N$ to be
sent to infinity at the end of the calculation. The essence of our
regularization method is to maintain all the crucial relations satisfied by
$R,T,v,w,\kappa_{o},\kappa_{e}$ but with finite norm for a modified $w$ as
long as $N$ is finite. The regularized automatically theory resolves the
associativity anomaly.

\subsection{Projecting out the anomalous sector}

We start from the example which we first explained in the last section. We
denote the mode space spanned by the basis $\cos(n\sigma)$ for $n=$odd (resp.
$n=$even) as $\mathcal{H}_{odd}$ (resp. $\mathcal{H}_{even})$. The matrices
$T$ and $R$ act on the mode spaces as
\begin{equation}
T\,:\,\mathcal{H}_{odd}\rightarrow\mathcal{H}_{even}\qquad R\,:\,\mathcal{H}%
_{even}\rightarrow\mathcal{H}_{odd},
\end{equation}
and they are the inverse of each other. We have discussed that the existence
of an eigenvector $v$ with zero eigenvalue implies the associativity anomaly
as $v=(RT)v\neq R(Tv)=0$. From the mathematical viewpoint, such an anomaly
should disappear in a sector with some restriction on the spaces
$\mathcal{H}_{even,odd}$ . Such a sector of string fields would presumably be
identified with the open string sector.

One natural restriction is the limitation of the elements of $\mathcal{H}$ to
square normalizable states. This restriction, however, is not enough to
guarantee associativity as seen in the case of Eq.(\ref{calV}) that has a
finite norm $\mathcal{V}^{\left(  o\right)  }$. Obviously the normalizable
vector $v\in\mathcal{H}_{odd}$ breaks associativity. We therefore proceed to
project it out from $\mathcal{H}_{odd}$ by using the projector,
\begin{equation}
P=1-v\bar{v}=\bar{T}T.
\end{equation}
We limit $\mathcal{H}_{odd}$ by using this projector $\mathcal{H}%
_{odd}^{\prime}=P\mathcal{H}_{odd}$ and redefine the operators in the
surviving subspace
\begin{equation}
T^{\prime}\equiv TP,\quad R^{\prime}\equiv PR.
\end{equation}
By using the identity, $R=\bar{T}+v\bar{w}$ in Eq.(\ref{tr}) one may easily
observe,
\begin{equation}
R^{\prime}T^{\prime}=P,\quad T^{\prime}R^{\prime}=1,\quad\mbox{and}\quad
R^{\prime}=\bar{T^{\prime}}.
\end{equation}
In a sense, $T^{\prime}$ and $R^{\prime}$ define the partial isometry between
$\mathcal{H}_{even}$ and $\mathcal{H}_{odd}^{\prime}$ and they become the
transpose of each other in the restricted space.

One subtlety is that there is naively a vector $w$ in $\mathcal{H}_{even}$
which causes a problem since $\bar{T}^{\prime}w=P\bar{T}w=Pv=0$ which seems to
imply the existence of a problematic zero eigenvalue. However, we note that we
restrict $\mathcal{H}$ to be square normalizable, and therefore the vector $w$
does not belong to $\mathcal{H}_{even}$ in this sense.

A cost for using this prescription is that we lose some basic properties of
$T$ and $R$ (\ref{eig1}--\ref{eig2}) after we project out the Hilbert spaces.
In particular, $\kappa_{o}$ should be replaced by a non-diagonal matrix
$P\kappa_{o} P$. In fact, the relations (\ref{eig1}--\ref{eig2}) are quite
singular since they imply that different sets of eigenvalues are related by
unitary transformations (as observed in \cite{r-MT})\footnote{We emphasize
however that these are not really unitary transformations when the subtleties
of the double sums are taken into account. Therefore, there really is no
contradiction.}. In this sense, losing these identities after we properly
define the space is natural. The failure of these identities is not desirable
since this would create some problems in the construction of the Virasoro
operators. Nevertheless one must also face the issue of anomalies that are in
conflict with the basic gauge symmetry of the action. We will come back to
this problem in our future work.

%\subsection{Implication to the open string Hilbert space}
%
%
%
We can interpret our constraint $\mathcal{H}_{odd}^{\prime}$ in terms of open
strings. When we take the Dirichlet boundary condition at the midpoint (odd
split string modes), we encountered a redundancy in the split string degree of
freedom involving $\left(  l^{\left(  o\right)  }+r^{\left(  o\right)
}\right)  \propto v$ and $\bar{x}$. We note that $\bar{x}$ is physically
essential to describe the vertex operator of the free boson which is the
exponential of $x^{\mu}\left(  \sigma\right)  $. While it may be possible to
remove the $\bar{x}$ variable, this reasoning suggests that it may not be a
good idea to proceed in this direction. So we take the other option, namely
projecting away the component of $\left(  l^{\left(  o\right)  }+r^{\left(
o\right)  }\right)  \propto v$ by applying the projector $P$. This
prescription is obviously consistent with our analysis.

When we use the Neumann boundary condition at the midpoint, the split string
variable is described by $\mathcal{H}_{even}$ and we do not need to consider
the projector for this case.

Some years ago, Strominger \cite{r-Strominger2} classified the inner
derivation of the open string Hilbert space into three subclasses
$\mathcal{O}$, $\mathcal{C}$, $\mathcal{I}$. The first one, $\mathcal{O}$, is
the inner derivative with respect to the open string field in a narrow sense
and the star product in this category is always associative. The second
category, $\mathcal{C}$, describes the variation of the closed string
background written in terms of the open string variable. He showed that the
element belonging to this subspace breaks associativity. The associator for
the closed string field then belongs to the third class $\mathcal{I}$ which is
described by the midpoint insertion of the primary field. The elements in
$\mathcal{I}$ commute with all the elements of the inner derivative.

This scenario can be applied to our simple situation. The inner product of the
open string is now represented by the commutators of the big matrices
described by the split string variables or by commutators involving the Moyal
star product. We have seen that associativity can be broken by the string
field degree of freedom that generates $\left(  P_{L}^{\mu}+P_{R}^{\mu
}\right)  $ that is related to the uniform translation of the open string (in
the Moyal basis this is the string field $A=\sum_{n,m}p_{2m}^{\mu}%
T_{2m,2n-1}v_{2n-1}\,$ under commutation, as seen above). In Strominger's
classification, this represents a single element in $\mathcal{C}$. We have
seen only one element since we considered only the algebra of string fields
linear in the modes $x^{\mu},p^{\mu}$. For the nonlinear string
configurations, the projection of the Hilbert space becomes more complicated
and we get more and more elements which belong to $\mathcal{C}$. It is not
easy to find a projection prescription to separate these configurations into
open/closed sectors. Therefore, we will resort to the regularized theory given
below which treats the issue of anomalies in a different manner.

\subsection{Subtlety of the vertex operators}

As we have seen, following Strominger's interpretation, the open string sector
can be identified by imposing certain constraints. The constraints can be
described in terms of the continuous variables $l(\sigma)$ and $r(\sigma)$ for
which constant shifts are allowed only in the opposite directions
$l(\sigma)+\varepsilon$ and $r(\sigma)-\varepsilon.$ More precisely the
allowed constant shifts are described by a kink at the midpoint and a
translation of the midpoint as discussed in Eqs.(\ref{ab1},\ref{ab2})
\begin{equation}
\delta x^{\mu}(\sigma)\propto b^{\mu}\,\left(  1+sign(\frac{\pi}{2}%
-\sigma)\right)  \label{allowed}%
\end{equation}
with a periodic sign function. This mode should be treated rather carefully.

This fact is relevant in recent developments in vacuum string field theory in
two contexts namely (i) the open string coupling to the closed string vertex
operator \cite{r-HI, GRSZ} and (ii) the ghost kinetic term as proposed by
\cite{GRSZ} namely fermionic ghost insertion at the midpoint.

In the first context, we recall that the proposed coupling of the open string
to the closed string background is,
\begin{equation}
\mathcal{O}_{V}=\int V(\pi/2)\Psi. \label{csc}%
\end{equation}
Here $V(\pi/2)$ is the midpoint insertion of the closed string vertex operator
$V$ acting on the open string field $\Psi$. The simplest choice for such
vertex operator is the closed string tachyon vertex $\exp(ik_{\mu}x^{\mu
}\left(  \pi/2\right)  )$. In the second context, the kinetic term of the VSFT
was proposed as,
\begin{equation}
S\sim\int\Psi\star(c(\pi/2)\Psi). \label{vsft}%
\end{equation}
If we represent the ghost field in terms of the free boson field $\phi(z)$
(identified as a 27$^{th}$ dimension $\phi\left(  \sigma\right)
=x_{27}\left(  \sigma\right)  ),$ then the $c(\pi/2)$ insertion is again
written in the form of a vertex operator $\exp(i\phi(\pi/2))$.

In the following, we show that the midpoint insertion of the vertex operator
discussed here can be precisely written in terms of the allowed kink
configuration which we have just mentioned, and that a deviation from the
midpoint kink is likely to create problems with associativity.

We take the fermionic ghost as the example. We consider its action at an
arbitrary point $\sigma_{0}.$ The ghost field $c^{\pm}(\sigma_{0})$ acts on
the string field in the 27$^{th}$ direction by creating a kink at $\sigma_{0}$
(see Eq.(3.41) of \cite{r-GT})
\begin{equation}
c^{\pm}(\sigma_{0})\Psi\lbrack\phi(\sigma)]=Ke^{i\epsilon(\sigma_{0})\frac
{\pi}{4}}e^{i\phi(\sigma_{0})}\Psi\lbrack\phi(\sigma)\pm\pi\theta(\sigma
_{0}-\sigma)].
\end{equation}
where the dependence on the other $26$ dimensions is suppressed.

For generic $\sigma_{0}\neq\pi/2$, the Fourier coefficients of the periodic
shift $\theta(\sigma_{0}-\sigma)$ are given by%
\begin{equation}
\theta\left(  \sigma_{0}-\sigma\right)  =\frac{\sigma_{0}}{\pi}+\sqrt{2}%
\sum_{n=1}^{\infty}\frac{\sqrt{2}\sin n\sigma_{0}}{n\pi}\cos\left(
n\sigma\right)
\end{equation}
The midpoint coordinate $\theta(\sigma_{0}-\frac{\pi}{2})=\bar{\theta}$ is
\begin{equation}
\bar{\theta}=\left\{
\begin{array}
[c]{cl}%
0\quad\quad & 0\leq\sigma_{0}<\pi/2\\
1\quad\quad & \pi/2<\sigma_{0}\leq\pi
\end{array}
\right.
\end{equation}
An expansion of $\theta\left(  \sigma_{0}-\sigma\right)  $ in terms of split
string modes can be developed as in sections (2.1) and (2.2) (odd modes). For
odd split string modes (Dirichlet at midpoint) the corresponding coefficients
are given as follows.
\begin{align}
&  \theta_{2n-1}^{\left(  l\right)  }=\frac{2\sqrt{2}}{\pi}\frac{\sin(\left(
2n-1\right)  \sigma_{0})}{2n-1},\quad\theta_{2n-1}^{\left(  r\right)
}=0,\qquad\mbox{for}\quad0\leq\sigma_{0}<\pi/2\label{leftmode}\\
&  \theta_{2n-1}^{\left(  r\right)  }=-\frac{2\sqrt{2}}{\pi}\frac{\sin(\left(
2n-1\right)  \sigma_{0})}{2n-1},\quad\theta_{2n-1}^{\left(  l\right)
}=0,\qquad\mbox{for}\quad\pi/2<\sigma_{0}\leq\pi/2
\end{align}
An expansion in terms of even split string modes can also be easily obtained
(Neumann at midpoint). We note that these odd split string mode expansion
coefficients have nonvanishing inner product with $v$. For example the
coefficients in (\ref{leftmode}) satisfies
\begin{equation}
\bar{v}\cdot\theta^{(l)}=2\sigma_{0}/\pi.
\end{equation}
This implies that the translation created by the kink $\theta\left(
\sigma_{0}-\sigma\right)  $ has a mode along the vector $\mathcal{V}^{\left(
o\right)  }$ of Eq.(\ref{calV}). As we have already discussed, this is an
extra mode which is at the very origin of the associativity anomaly.

Let us now consider the ghost very special point $\sigma_{0}=\pi/2.$ The kink
now creates a translation proportional to the periodic function $\theta\left(
\frac{\pi}{2}-\sigma\right)  $ which is given in terms of the periodic sign
function by%
\begin{equation}
\theta(\frac{\pi}{2}-\sigma)=\frac{1}{2}+\frac{1}{2}sign\left(  \frac{\pi}%
{2}-\sigma\right)
\end{equation}
This is an allowed translation as in Eq.(\ref{allowed}). $.$Indeed the mode
expansion for this function is given in terms of the full string modes, odd
split string modes, and even split string modes as follows
\begin{align}
&  \theta\left(  \frac{\pi}{2}-\sigma\right) \nonumber\\
&  =\frac{1}{2}+\sqrt{2}\sum_{n=1}^{\infty}\frac{v_{2n-1}}{2}\cos\left(
\left(  2n-1\right)  \sigma\right) \label{thetafull}\\
&  =%
\begin{tabular}
[c]{|c|c|c|}\hline
left/right & odd split modes $\,$ & even split modes\\\hline
midpoint & $\frac{1}{2}$ & $\frac{1}{2}$\\\hline
left modes & $\frac{1}{2}v_{2n-1}$ & $\frac{1}{2\left(  1+\bar{w}w\right)
}w_{2m}$\\\hline
right modes & $-\frac{1}{2}v_{2n-1}$ & $-\frac{1}{2\left(  1+\bar{w}w\right)
}w_{2m}$\\\hline
\end{tabular}
\label{thetasplit}%
\end{align}
In particular, the odd split string modes are given by the first column in
this table
\begin{equation}
\bar{\theta}=1/2,\quad\theta_{2n-1}^{\left(  l\right)  }=-\theta
_{2n-1}^{\left(  r\right)  }=\frac{1}{2}v_{2n-1}.
\end{equation}
which shows again in detail that it is of the allowed type, as in
Eqs.(\ref{ab1},\ref{ab2}). Thus, we see that except for the case, $\sigma
_{0}=\pi/2$, the kink created by the vertex is anomalous. In this respect, the
construction of the vertex operators is a very subtle problem.

%Thus the midpoint ghost generates a translation of the allowed form on the
%left/right parts of $\phi (\sigma )$ but not in $\cH'$.
%Namely, the effect of the ghost at the
%midpoint $\sigma _{0}=\frac{\pi }{2}$ is to create a kink which is
%compatible with the open string sector.
%
%
%

A lesson that we can draw here may be the following. The guiding principle to
determine the closed string coupling (\ref{csc}) or the kinetic term
(\ref{vsft}) was that they enjoyed an enhanced symmetry if inserted at the
midpoint. From our point of view, the midpoint is the safest choice because
associativity is preserved. However, the regularization offered to define them
do not seem to enjoy the same properties.

\section{Consistent regularization}

The anomaly occurred because of the infinite norm of the vector $w.$ To
analyze the anomaly more carefully it is necessary to introduce a consistent
regulator. This can be done by formulating a cutoff version of the theory
using finite rank matrices which truncate the theory to a finite number of
oscillator modes $\left(  x_{2n},p_{2n}\right)  $ with $n=1,2,\cdots,N.$ Such
a cutoff is desirable more generally to regulate string field theory. It could
also be useful for numerical estimates.

In this section we will denote the finite matrices with the same symbols as
the infinite matrices. Thus, we have square $N\times N$ matrices
$T,R,\kappa_{e,}\kappa_{o}$ and $N$ dimensional column matrices $v,w.$ These
finite matrices may depend on the cutoff $N$ not only through their rank but
also explicitly in the matrix elements. We will see that except for the a
general structure that we will explain, we seldom need the details of the $N$
dependence, for which there is a certain amount of flexibility. In any case,
the finite matrices become the infinite matrices we discussed before when
$N\rightarrow\infty.$

To have a consistent theory, the finite rank matrices must obey the same
relations among themselves that were obeyed by the infinite rank matrices. The
list of all the relations that must be satisfied for arbitrary finite $N$ are
given in Eqs.(\ref{tr}-\ref{Rw}) excluding the hasty infinite limit in
Eq.(\ref{tt2}). These relations are satisfied by the infinite matrices. Here
we will present the general solution for $R,T,v,w,\kappa_{o},\kappa_{e}$ that
satisfy these relations at any $N.$

When $N\rightarrow\infty$ the sum $\bar{w}w$ diverges even though all
components of the vector are finite. The associativity anomaly arises from
this infinite norm. Associativity is restored by keeping track of this
divergence when multiple sums are involved, and sending $N\rightarrow\infty$
only at the end of a calculation. In taking the limit any explicit $N$
dependence in the matrix elements of the matrices $T,R,\kappa_{e,}\kappa_{o}%
$,$v,w$ should be taken into account. However, the important infinity usually
occurs in the form $\bar{w}w,$ therefore keeping track of this expression is
mainly what is needed in most cases to extract the unique values consistent
with associativity.

First we give the general solution for $R,T$ that satisfy all the relations
except for $R=\kappa_{o}^{-2}\bar{T}\kappa_{e}^{2}.$ This is given in terms of
a general orthogonal matrix $S$ as follows
\begin{equation}
T=\left(  1+w\bar{w}\right)  ^{-\frac{1}{2}}S,\quad R=\bar{S}\left(
1+w\bar{w}\right)  ^{\frac{1}{2}} \label{trs}%
\end{equation}
Expanding the square roots in a power series, one may write the matrices
$\left(  1+w\bar{w}\right)  ^{\pm\frac{1}{2}}$ in the form
\begin{equation}
\left(  1+w\bar{w}\right)  ^{\pm\frac{1}{2}}=1+\frac{1}{\bar{w}w}\left(
\left(  1+\bar{w}w\right)  ^{\pm\frac{1}{2}}-1\right)  w\bar{w}.
\end{equation}
Thus, for any orthogonal matrix $S$ that satisfies $S\bar{S}=1=\bar{S}S,$ and
any vector $w,$ one constructs
\begin{equation}
v=\bar{T}w=\frac{\bar{S}w}{\sqrt{1+\bar{w}w}} \label{vsw}%
\end{equation}
and then easily verify that all the relations hold. Other than leaving $S$
unspecified, this form of the solution for $T,R,v$ is unique for a given $w.$

Next we turn to the condition $R=\kappa_{o}^{-2}\bar{T}\kappa_{e}^{2}.$ By
inserting the $R,T$ given above this relation takes the form
\begin{equation}
\kappa_{o}^{2}=\bar{T}\kappa_{e}^{2}T=\bar{S}\left(  1+w\bar{w}\right)
^{-\frac{1}{2}}\kappa_{e}^{2}\left(  1+w\bar{w}\right)  ^{-\frac{1}{2}}S
\label{wowe}%
\end{equation}
By a linear transformation of $S$ and $w,$ one may always go to a basis in
which both spectrum matrices $\kappa_{o},\kappa_{e}$ are diagonal. Thus,
without loosing generality we assume diagonal $\kappa_{o},\kappa_{e}.$ In such
a basis we see that the meaning of $S$ in the above equation is that it is the
orthogonal matrix that diagonalize the symmetric matrix $\left(  1+w\bar
{w}\right)  ^{-\frac{1}{2}}\kappa_{e}^{2}\left(  1+w\bar{w}\right)
^{-\frac{1}{2}}.$

Let us now determine the spectrum $\kappa_{o}$ by solving the eigenvalue
condition for the matrix $\left(  1+w\bar{w}\right)  ^{-\frac{1}{2}}\kappa
_{e}^{2}\left(  1+w\bar{w}\right)  ^{-\frac{1}{2}}.$ We wish to solve the
secular equation
\begin{equation}
\det\left(  \left(  1+w\bar{w}\right)  ^{-\frac{1}{2}}\kappa_{e}^{2}\left(
1+w\bar{w}\right)  ^{-\frac{1}{2}}-\lambda\right)  =0. \label{sec}%
\end{equation}
The determinant can be computed as follows
\begin{align}
&  \det\left(  \left(  1+w\bar{w}\right)  ^{-\frac{1}{2}}\left(  \kappa
_{e}^{2}-\lambda\left(  1+w\bar{w}\right)  \right)  \left(  1+w\bar{w}\right)
^{-\frac{1}{2}}\right) \\
&  =\det\left(  \left(  1+w\bar{w}\right)  ^{-\frac{1}{2}}\right)  ^{2}%
\,\det\left(  \kappa_{e}^{2}-\lambda\left(  1+w\bar{w}\right)  \right) \\
&  =\frac{\det\left(  \kappa_{e}^{2}-\lambda\right)  }{1+\bar{w}w}\det\left(
1-\frac{\lambda}{\kappa_{e}^{2}-\lambda}w\bar{w}\right) \\
&  =\frac{\det\left(  \kappa_{e}^{2}-\lambda\right)  }{1+\bar{w}w}\left(
1-\bar{w}\frac{\lambda}{\kappa_{e}^{2}-\lambda}w\right)
\end{align}
The only way to have a vanishing determinant is by the vanishing of the last
factor, therefore the secular equation becomes
\begin{equation}
\frac{1}{\lambda}+\sum_{n=1}^{N}\frac{w_{2n}^{2}}{\lambda-\kappa_{2n}^{2}}=0.
\label{eigen}%
\end{equation}
This equation can be rewritten as an $N^{th}$ order polynomial in $\lambda,$
and therefore it has $N$ roots for $\lambda$. The $N$ roots correspond to the
diagonal matrix $\kappa_{o}.$ Therefore we have the following $N$ relations
among the $3N$ numbers $w_{2n},\kappa_{2n},\kappa_{2n-1}$%
\begin{equation}
\frac{1}{\kappa_{2m-1}^{2}}+\sum_{n=1}^{N}\frac{w_{2n}^{2}}{\kappa_{2m-1}%
^{2}-\kappa_{2n}^{2}}=0. \label{Nrelations}%
\end{equation}
for $m=1,2,\cdots,N.$ Recall that the meaning of the eigenvalues $\kappa
_{2n},\kappa_{2n-1}$ is that they represent the frequencies of oscillations of
the string modes, while $w_{2n}$ is related to the modes that determine the
difference between the center of mass point and the midpoint $x_{0}-\bar{x}$.
These $N$ equations determine uniquely the $w_{2n}^{2}$ for arbitrary
$\kappa_{2n},\kappa_{2n-1}.$ The unique solution is
\begin{equation}
w_{2n}^{2}\left(  N\right)  =\sum_{m=1}^{N}\left(  M^{-1}\right)
_{2n,2m-1}\frac{1}{\kappa_{2m-1}^{2}} \label{w2n}%
\end{equation}
where $\left(  M^{-1}\right)  _{2n,2m-1}$ is the inverse of the $N\times N$
matrix determined by the frequencies
\[
M_{2m-1,2n}=\frac{-1}{\kappa_{2m-1}^{2}-\kappa_{2n}^{2}},
\]
We expect that in the large $N$ limit $w_{2n}=\sqrt{2}\left(  -1\right)
^{n+1}$ when $\kappa_{2n}=2n$ and $\kappa_{2m-1}=2m-1.$ Indeed it is easily
verified that
\begin{equation}
\frac{1}{\left(  2m-1\right)  ^{2}}+\sum_{n=1}^{\infty}\frac{2}{\left(
2m-1\right)  ^{2}-\left(  2n\right)  ^{2}}=0 \label{iden}%
\end{equation}
is satisfied for every integer $m,$ showing that we have the expected solution
in the large $N$ limit.

At finite $N,$ we have the freedom to choose freely $2N$ numbers $\kappa
_{2n}\left(  N\right)  ,\kappa_{2n-1}\left(  N\right)  $ and determine the $N$
numbers $w_{2n}^{2}\left(  N\right)  $ from Eq.(\ref{w2n}). Symmetry
considerations may dictate a particular pattern for the $N$ dependence of
$\kappa_{2n}\left(  N\right)  ,\kappa_{2n-1}\left(  N\right)  $ at some stage
of our investigation. For now, as an example, suppose we make the choice
$\kappa_{2n}=2n$ and $\kappa_{2n}=2n-1$ just like at infinite $N,$ and then
determine $w_{2n}^{2}\left(  N\right)  $ as a function of $N$. The solutions
can be obtained numerically and their dependence on $N$ can be studied. Also,
the unique diagonalizing matrix $S$ can be obtained numerically and its
dependence on $N$ can be studied. Some examples for $N=2,3,10$ are given below
as follows from Eq.(\ref{w2n}), they show consistency with $w_{2n}=\sqrt
{2}\left(  -1\right)  ^{n+1}$ as $N$ gets larger
\begin{align}
\left.  w_{2n}^{2}\right|  _{N=2}  &  =\left(
\begin{array}
[c]{cc}%
\frac{1}{3} & \frac{1}{15}\\
-\frac{1}{5} & \frac{1}{7}%
\end{array}
\right)  ^{-1}\left(
\begin{array}
[c]{c}%
1\\
\frac{1}{9}%
\end{array}
\right)  =\left(
\begin{array}
[c]{c}%
\frac{20}{9}\\
\frac{35}{9}%
\end{array}
\right)  =\left(
\begin{array}
[c]{c}%
2.\,\allowbreak222\\
3.\,\allowbreak888
\end{array}
\right) \\
\left.  w_{2n}^{2}\right|  _{N=3}  &  =\left(
\begin{array}
[c]{ccc}%
\frac{1}{3} & \frac{1}{15} & \frac{1}{35}\\
-\frac{1}{5} & \frac{1}{7} & \frac{1}{27}\\
-\frac{1}{21} & -\frac{1}{9} & \frac{1}{11}%
\end{array}
\right)  ^{-1}\left(
\begin{array}
[c]{c}%
1\\
\frac{1}{9}\\
\frac{1}{25}%
\end{array}
\right)  =\left(
\begin{array}
[c]{c}%
\frac{21}{10}\\
\frac{63}{25}\\
\frac{231}{50}%
\end{array}
\right)  =\left(
\begin{array}
[c]{c}%
2.\,1\\
2.\,52\\
4.\,62
\end{array}
\right)
\end{align}
For $N=10$, we get the 10 numbers
\[
w_{2n}^{2}%
=(2.\,009,\ 2.\,039,\ 2.\,091,\ 2.\,171,\ 2.\,290,\ 2.\,465,\ 2.\,734,\ 3.\,190,\ 4.\,141,\ 8.\,076\ )
\]
which approaches $w_{2n}^{2}=2$ except for the last few components. For faster
convergence one may take advantage of the freedom in the choices we can make
freely in the $N$ dependence of $\kappa_{2n}\left(  N\right)  ,\kappa
_{2n-1}\left(  N\right)  .$ However, we will not exercise this choice until we
determine whether some symmetry considerations dictate a specific $N$ dependence.

There are some relations among the $w_{2n}^{2}\left(  N\right)  ,$
$\kappa_{2n}\left(  N\right)  ,$ $\kappa_{2n-1}\left(  N\right)  $ which can
be read off directly from Eq.(\ref{eigen}) without knowing explicitly the
$\kappa_{2n}\left(  N\right)  ,\kappa_{2n-1}\left(  N\right)  $. We first
rewrite Eq.(\ref{eigen}) by multiplying it with $\lambda\prod_{n=1}%
^{N}(\lambda-\kappa_{2n}^{2})$ and expressing the secular determinant it terms
of its eigenvalues as in the right hand side below
\begin{align}
&  \prod_{n=1}^{N}(\lambda-\kappa_{2n}^{2})+\sum_{n=1}^{N}\bar{w}_{2n}%
w_{2n}\lambda\prod_{i(\neq n)}(\lambda-\kappa_{2i}^{2})\nonumber\\
&  =(1+\bar{w}w)\prod_{n=1}^{N}(\lambda-\kappa_{2n-1}^{2})
\end{align}
The overall coefficient on the right hand side is determined by comparing the
highest power of $\lambda$ on both sides. The same relation follows from
Eq.(\ref{wowe}) after subtracting $\lambda$ from both sides and computing the
determinant. By comparing the coefficients of various powers of $\lambda$ on
both sides we can derive many relations. In particular, by comparing the
coefficient of the zeroth power one finds
\begin{equation}
\det\left(  \frac{\kappa_{e}^{2}}{\kappa_{o}^{2}}\right)  =\prod_{n=1}%
^{N}\frac{\kappa_{2n}^{2}}{\kappa_{2n-1}^{2}}=1+\bar{w}w.
\end{equation}
The first power in $\lambda$ gives
\begin{equation}
\sum_{i=1}^{N}\frac{1+w_{2i}^{2}}{\kappa_{2i}^{2}}=\sum_{i=1}^{N}\frac
{1}{\kappa_{2i-1}^{2}}%
\end{equation}
and the $k$'th power in $\lambda$ yields
\begin{equation}
\sum_{i_{1}\neq i_{2}\cdots\neq i_{k}=1}^{N}\frac{1+w_{2i_{1}}^{2}+w_{2i_{2}%
}^{2}+\cdots+w_{2i_{k}}^{2}}{\kappa_{2i_{1}}^{2}\kappa_{2i_{2}}^{2}%
\cdots\kappa_{2i_{k}}^{2}}=\sum_{i_{1}\neq i_{2}\cdots\neq i_{k}=1}^{N}%
\frac{1}{\kappa_{2i_{1}-1}^{2}\kappa_{2i_{2}-1}^{2}\cdots\kappa_{2i_{k}-1}%
^{2}}%
\end{equation}

In particular, the ratio $\det\left(  \kappa_{e}^{2}/\kappa_{o}^{2}\right)
=1+\bar{w}w$ which was computed from the zeroth power in $\lambda$ has a
universal form in terms of $w$ independent of the specific choice of
$\kappa_{e},\kappa_{o}$. Note that in the infinite $N$ limit $1+\bar{w}w$ is
the periodic delta function $\delta\left(  \sigma-\frac{\pi}{2}\right)  $with
vanishing argument, as seen from Eq.(\ref{delta})
\begin{equation}
1+\bar{w}w=\pi\delta\left(  0\right)  .
\end{equation}
This relation also implies that the determinant of $T$, $\bar{T}$ can be
computed from Eq.(\ref{tt},\ref{wowe})
\begin{equation}
\det(\bar{T}T)=\det(\kappa_{o}^{2}/\kappa_{e}^{2})=(1+\bar{w}w)^{-1}.
\end{equation}
By replacing this in Eqs.(\ref{tr},\ref{trs}) we learn
\begin{equation}
\left(  \det R\right)  ^{-1}=\det T=\det\left(  \frac{\kappa_{o}}{\kappa_{e}%
}\right)  =(1+\bar{w}w)^{-1/2}.
\end{equation}
If we use the choice $w_{2n}=\sqrt{2}(-1)^{n+1}$ at large $N,$ the right hand
side vanishes as $O(1/\sqrt{N})$ in the large $N$ limit.

Having satisfied the crucial relations for $R,T,v,w,\kappa_{o},\kappa_{e}$ at
finite $N,$ we may next easily represent some cutoff versions of distributions
such as the theta or delta functions given in Eqs.(\ref{delta},\ref{thetafull}%
,\ref{thetasplit}) in which $v,w$ appear as fundamental entities.

The above are examples of expressions and relations that could not be uniquely
determined without a consistent cutoff. These results are useful in explicit
computations that will be presented in \cite{r-BarsMatsuo}.

\section{Conclusion and Outlook}

In this paper, we presented fundamental aspects of the associativity anomaly
in Witten's string field theory, and indicated that it arises from the
matrices that map the open string variable to the split string variable. We
argued that this anomaly does not come from the peculiarity of using the split
string variables but can be related to the midpoint issues noticed in the
literature in other contexts. We proposed some prescriptions to deal with the
anomaly. One was based on an attempt to separate open/closed string modes in
the presence of an infinite number of modes, the other was based on a
systematic cutoff version of the theory.

We have to emphasize that our arguments are not yet complete. For example, in
the first prescription, although we may identify some anomaly causing singular
string fields that are linear, the vertex operators $V_{N}$ \cite{r-GJ} and
other string field configurations are expected to contain nonlinear singular
fields associated with closed strings . In this sense, it is not completely
clear how to generalize the open/closed separation systematically to all
string fields.

On the other hand, in the regularization with a finite number of
modes, we do not have such a difficulty. We pointed out that there
remains some arbitrariness in fixing the matrices $\kappa_{e}$ and
$\kappa_{o}$ where we have not yet found a principle to determine
them. A related issue is how to represent conformal symmetry with
a finite number of modes. One idea which we have not explored yet
is to use quantum groups, which may lead to a determination of
$\kappa_{e}$ and $\kappa_{o}$ consistently with conformal
symmetry. Nevertheless, our prescription is successful in
providing a systematic regulator.

After providing a regularization as in this paper, the Moyal approach
\cite{r-Bars} gives a very simple framework to calculate various quantities in
string field theory. In a forthcoming paper, we will discuss explicit
computations, including a discussion of the Virasoro generators in terms of
Moyal variables (in the $N\rightarrow\infty$ limit), and calculations of
off-shell $n$-point amplitudes.

\bigskip

\begin{center}
\noindent{\large \textbf{Acknowledgments}}
\end{center}

I.B. is supported in part by a DOE grant DE-FG03-84ER40168. Y.M. is supported
in part by Grant-in-Aid (\# 13640267) and in part by Grant-in-Aid (\# 707)
from the Ministry of Education, Science, Sports and Culture of Japan. Both
Y.M. and I.B. thank the NSF and the JSPS for making possible the collaboration
between Tokyo University and USC through the collaborative grants, NSF-9724831
and JSPS (US-Japan cooperative science program). I.B. is grateful to Tokyo
University and Y.M. is grateful to the Caltech-USC Center for hospitality.

\end{document}